\documentclass[runningheads]{llncs}

\usepackage{eccv}

\usepackage{eccvabbrv}

\usepackage{graphicx}
\usepackage{booktabs}
\usepackage{multirow}
\usepackage{bm}
\usepackage{colortbl}
\usepackage{algorithm2e, setspace}

\usepackage[accsupp]{axessibility}  
    
\newcommand{\myparagraph}[1]{\noindent\textbf{#1}}

\let\titleold\title
\renewcommand{\title}[1]{\titleold{#1}\newcommand{\thetitle}{#1}}

\usepackage{hyperref}

\usepackage{orcidlink}

\begin{document}

\title{DualDn: Dual-domain Denoising via Differentiable ISP} 

\titlerunning{DualDn}

\author{Ruikang Li\inst{1,2}\thanks{This work was done during Ruikang Li's internship at Shanghai Artificial Intelligence Laboratory.} \and
Yujin Wang\inst{1}\textsuperscript{\dag} \and
Shiqi Chen\inst{3} \and 
Fan Zhang\inst{1} \and
Jinwei Gu\inst{2} \and
Tianfan Xue\inst{2}
}
\renewcommand{\thefootnote}{\fnsymbol{footnote}}
\footnotetext[0]{\hspace{-1.1em} \textsuperscript{\dag} Corresponding author.}

\authorrunning{R.~Li et al.}

\institute{Shanghai Artificial Intelligence Laboratory \and
The Chinese University of Hong Kong\\ \and
Zhejiang University\\
}

\maketitle

\begin{abstract}
Image denoising is a critical component in a camera's Image Signal Processing (ISP) pipeline. There are two typical ways to inject a denoiser into the ISP pipeline: applying a denoiser directly to captured raw frames (raw domain) or to the ISP's output sRGB images (sRGB domain). However, both approaches have their limitations. Residual noise from raw-domain denoising can be amplified by the subsequent ISP processing, and the sRGB domain struggles to handle spatially varying noise since it only sees noise distorted by the ISP. Consequently, most raw or sRGB domain denoising works only for specific noise distributions and ISP configurations. To address these challenges, we propose \textbf{DualDn}, a novel learning-based dual-domain denoising.
Unlike previous single-domain denoising, DualDn consists of two denoising networks: one in the raw domain and one in the sRGB domain. The raw domain denoising adapts to sensor-specific noise as well as spatially varying noise levels, while the sRGB domain denoising adapts to ISP variations and removes residual noise amplified by the ISP. Both denoising networks are connected with a differentiable ISP, which is trained end-to-end and discarded during the inference stage.
With this design, DualDn achieves greater generalizability compared to most learning-based denoising methods, as it can adapt to different unseen noises, ISP parameters, and even novel ISP pipelines. Experiments show that DualDn achieves state-of-the-art performance and can adapt to different denoising architectures. Moreover, DualDn can be used as a plug-and-play denoising module with real cameras without retraining, and still demonstrate better performance than commercial on-camera denoising. The project website is available at:~\url{https://openimaginglab.github.io/DualDn/} 
\end{abstract}

\begin{figure}[t]
  \centering
    \includegraphics[width=\linewidth]{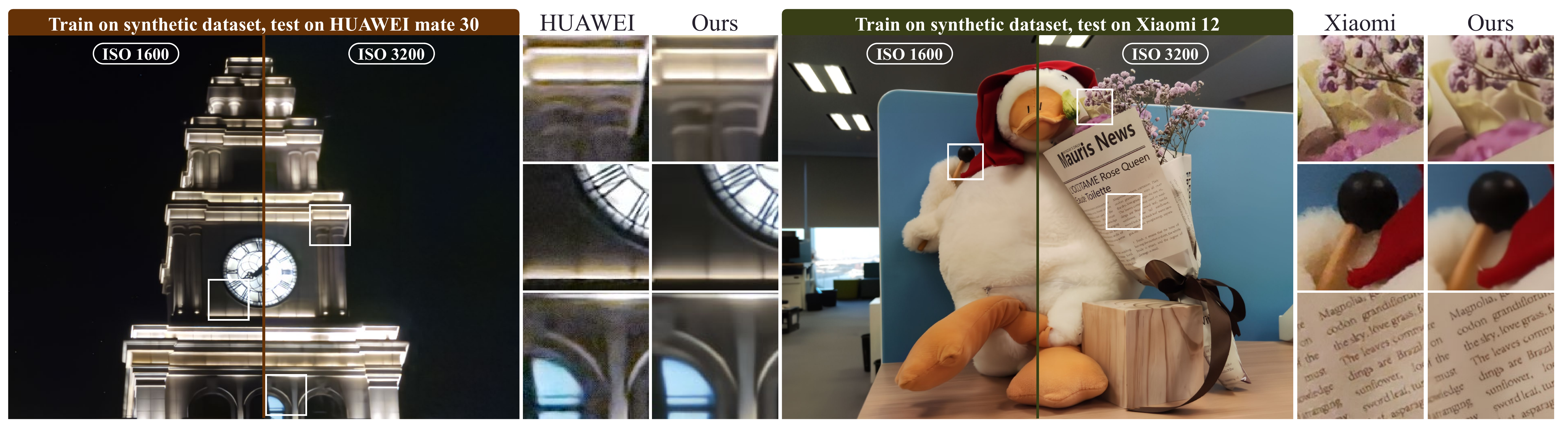}
    \caption{Illustrating the generalizability of our dual-domain denoising. The proposed denoising outperforms commercial denoising algorithms on two smartphone cameras under various ISOs. Notice that our denoising is only trained on synthetic images, without using any images from these cameras or their ISP pipelines during training.
    }
    \label{Fig.1: Visual results of denoising in different domains.}
\end{figure}

\section{Introduction}
\label{Sec.1: Introduction}

In recent years, neural network denoising algorithms~\cite{liang2021swinir,chen2022simple,zamir2022learning} have demonstrated spectacular quality on both real-captured images~\cite{abdelhamed2018high, plotz2017benchmarking, abdelhamed2020ntire, abdelhamed2019ntire} and synthetic ones~\cite{zhang2017beyond, zhang2021plug, wei2020physics, li2023ntire}. Typically, to train these denoising networks, a training set is first collected for a specific camera model with known noise levels. Networks trained on that dataset often work well for that setup, but the performance drops when applied to other camera models or different noise levels.
Researchers tried to mitigate this issue through generalizable training schedules~\cite{chen2023masked}, but denoising networks trained in this manner are either slow or perform worse than classical denoising networks, especially when dealing with severe noise. As a result, designing a universally adaptable denoising framework, applicable to various sensor noises and camera ISPs, remains a challenging task.

\begin{figure}[t]
  \centering
    \includegraphics[width=0.84\linewidth]{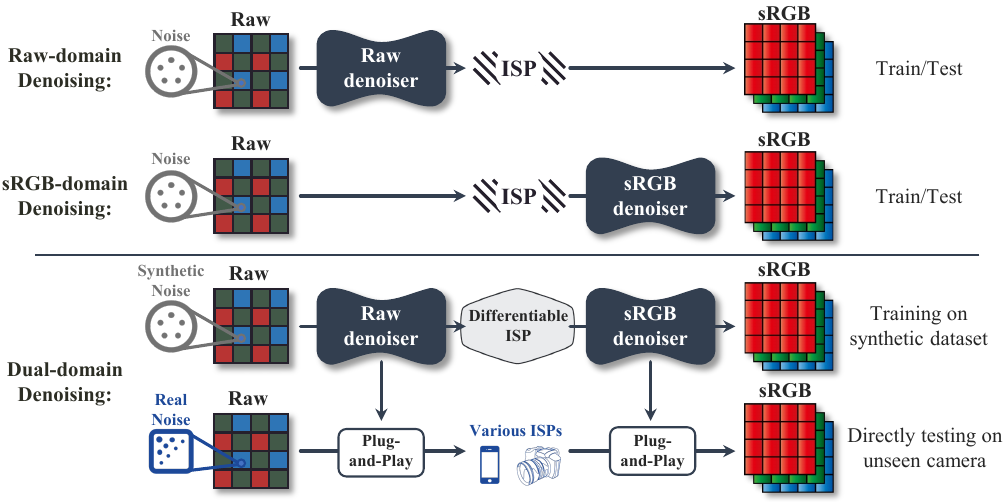}
    \caption{Compare single-domain and dual-domain denoising. Noise in raw domain is sensor-specific and ISO-dependent, and ISP is device-related and user-preferred. Only denoising in dual-domain can properly deal with noise and ISP variations respectively.}
    \label{fig:intro}
\end{figure}

Our key observation is that the limited generalization ability is related to the domain to which denoising algorithms are applied. As shown in~\cref{fig:intro}, in a typical camera pipeline, sensor noise is introduced in the raw domain and is propagated to the final sRGB images after the raw image is processed by the Image Signal Processor (ISP) pipeline. Therefore, there are two typical ways to apply denoising: in the raw domain and in the sRGB domain. Raw domain denoising often generalizes well to different noise distributions, as there are well-studied noise models that correlate noise distributions with a few parameters that can be obtained through calibration~\cite{jin2023lighting}. However, the minor residual noise or artifacts after raw-domain denoising may be magnified by ISP processing, resulting in sub-optimal final sRGB images (the first-row of~\cref{fig:intro}). For example, demosaicing often produces high-frequency artifacts around image edges,  and tone-mapping may amplify the residual noise in dark regions. Moreover, ISP processes are unknown to raw domain denoising, and thus their performance drops when switching to different ISP pipelines.

On the other hand, sRGB-domain denoising is directly applied to the final sRGB image generated by the ISP and thus can generalize to different ISP pipelines. Nonetheless, most sRGB-domain denoising networks are trained with a single noise distribution~\cite{zhang2021plug, jiang2022fast}. On the other side, the actual sRGB noise distribution is complex, mainly because the calibratable raw noise is distorted by ISP processing, making it unpredictable in the sRGB domain (the second-row of~\cref{fig:intro}). As a result, the performance of sRGB denoising declines when applied to strong noise either in low-light conditions or with spatially varying noise.

Therefore, in this work, to account for both sensor noise and ISP processing variations, we propose a novel dual-domain denoising pipeline that consists of a raw denoiser and an sRGB denoiser, as shown in the bottom part of~\cref{fig:intro}. Utilizing raw denoising allows the network to handle spatially varying noise and generalize to different sensors with the help of calibrated noise levels. After the denoised raw signal is processed by the ISP, the second sRGB denoiser further eliminates amplified and correlated residual noise introduced by the ISP. Compared to single-domain denoising, our dual-domain denoising produces visually pleasing images with fine details. To train these dual-domain denoising networks, we also propose a training scheme that includes augmented noise distribution and ISP parameters, as well as differentiable ISP pipelines that support the joint training of both denoisers simultaneously.

To the best of our knowledge, while traditional ISP pipelines often place noise filters in both the raw and sRGB domains~\cite{knaus2013dual,brown2019understanding}, we are the first to propose a learning-based dual-domain denoising framework that is end-to-end trainable. To achieve this, we use differentiable ISPs and noise map mechanisms as ``bridges''. Additionally, we generate different noises for the raw domain and different ISPs for the sRGB domain during training, enabling the raw and sRGB denoisers to jointly handle both noise and ISP variations. Moreover, our experimental results on both synthetic and real images show significant improvements over single-domain baselines, especially when applied to unseen ISPs or noise distributions. Furthermore, although our model is trained only on synthetic images, it can be applied to black-box ISPs in commercial smartphones and even outperforms the denoising algorithms carefully designed for their cameras, as shown in~\cref{Fig.1: Visual results of denoising in different domains.}.

\section{Related Work}
\label{Sec.2: Related Work}

Since deep networks were first used for image denoising in 2015~\cite{liang2015stacked,xu2015denoising}, they have achieved superior image quality compared to traditional methods~\cite{chan2005salt,buades2005non,aharon2006k,elad2006image,dabov2007image,mairal2009non,zoran2011learning}. This performance is due to extensive training data and advanced network architectures~\cite{tian2020deep,goyal2020image,yapici2021review,zhang2021designing,liang2021mutual}. Denoising networks, from ISP design perspective, are categorized into raw domain-based and sRGB domain-based methods.

\paragraph{Denoising in raw domain.}
It is straightforward to collect real noisy-clean raw pairs for training~\cite{chen2018learning,abdelhamed2018high}, but time-consuming for each type of image sensor. By taking advantage of the extensive studies on sensor noise~\cite{foi2007noise,foi2008practical,konnik2014high,zhu2016noise,zhang2017improved}, it is efficient to synthesize the training pairs based on a calibrated noise model~\cite{fu2023srgb}. More complicated noise models~\cite{kingma2018glow,henz2020synthesizing,abdelhamed2019noise}, elaborate calibrations~\cite{monakhova2022dancing}, and sophisticated fine-tuning strategy~\cite{jin2023lighting} have been proposed to improve the capability of a denoising network for a specific image sensor~\cite{wei2020physics,wei2021physics}. Nevertheless, they may fail to cover casual real-world cases due to diverse lighting conditions or varying ISP settings. Moreover, they are not capable of handling the complicated noise in the sRGB domain, as ISP processing significantly alters noise characteristics~\cite{fu2023srgb}.

\paragraph{Denoising in sRGB domain.}
Most sRGB-domain denoising networks~\cite{zhang2017beyond,zhang2018ffdnet,cheng2021nbnet,zhang2021plug,liang2021swinir,jiang2022fast,zamir2022restormer} assume that the noise is additive white Gaussian noise (AWGN). While AWGN is a simple and general model, it fails to accurately represent the noise characteristics in the sRGB domain~\cite{guo2019toward}. Recent approaches introduce random masks~\cite{chen2023masked} and controllable modules~\cite{zhang2023real} to make the denoising network more adaptable for real applications. However, when tested on unseen cameras, these methods often produce over-smoothed and unsatisfactory results.

\section{Method}
\label{sec:Method}

Given a raw image with wide-ranging noise from an unknown camera, our goal is to restore a clean sRGB image using trained denoising networks and a target ISP. Our denoising network is not tied to any specific ISP and only replaces its denoising modules. In this section, we first present the overall framework of our DualDn. Then we describe the design of each part of DualDn in detail.

\begin{figure}[t]
  \centering
    \includegraphics[width=\linewidth]{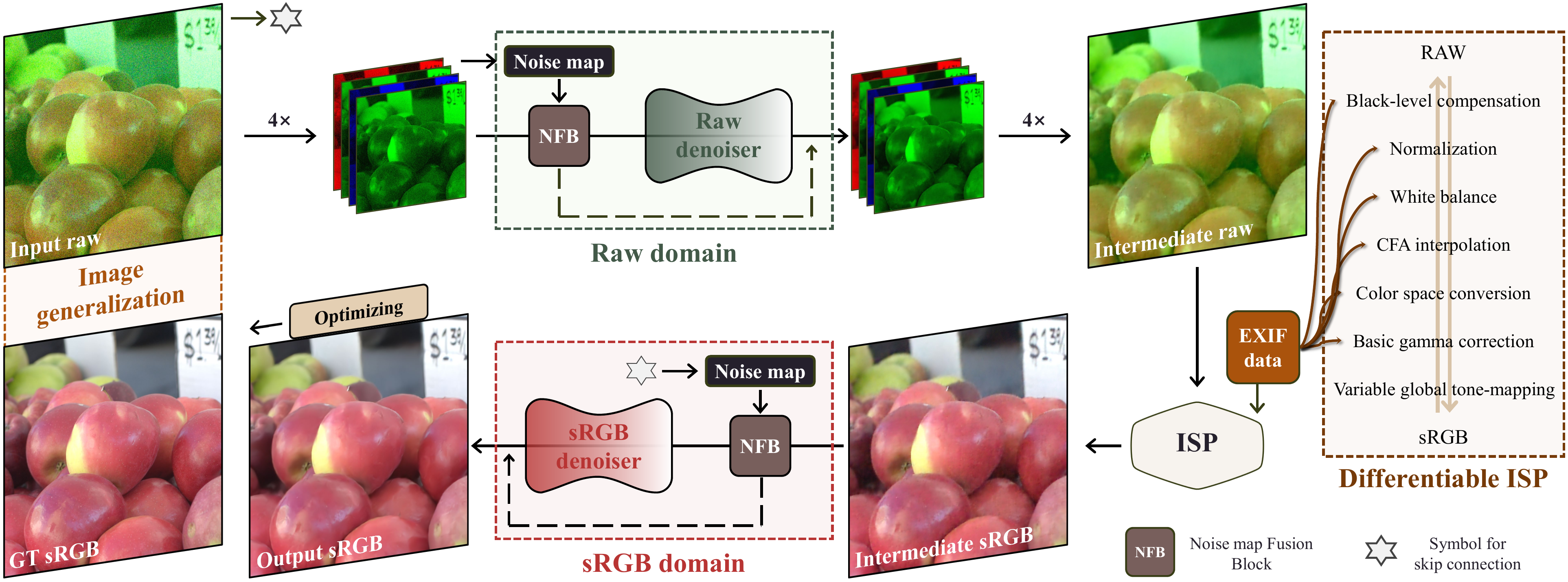}
    \caption{The overall pipeline of DualDn. It consists of 3 key components: (a) image generalization with various noise, (b) dual-domain denoising with noise map fusion, and (c) differentiable ISP with corresponding EXIF data and variable ISP parameters.}
    \label{fig:Our_pipeline}
\end{figure}

The overall pipeline of DualDn is shown in~\cref{fig:Our_pipeline}. Given a clean raw patch, we generate the corresponding clean sRGB patch according to its EXIF metadata and synthesize a noisy raw patch based on the noise model. The noisy raw patch is then processed through raw denoising and sRGB denoising to produce the final sRGB patch. This process adopts a differentiable ISP to simulate intermediate ISP processing between the raw and sRGB domains, utilizing EXIF metadata and keeping other parameters adjustable, as explained in~\cref{sec:Differentiable_ISP}. 

\subsection{Image Generalization}
\label{sec:image_generalization}

We synthesize noisy raw images from the clean ones using a noise model and ample sets of model parameters. To enhance the generalization of our method, we mainly consider shot noise and read noise similar to~\cite{brooks2019unprocessing,wei2021physics}, while ignoring device-specific noise such as fixed pattern noise~\cite{boukhayma2016ultra}. Nevertheless, the experiments in~\cref{sec:Real_world_Scenarios_Testing} indicate that training with this efficient noise model can generalize well to in-the-wild images from unseen cameras with unknown noise. 

The shot noise and the read noise conform to a Poisson and a Gaussian distribution~\cite{foi2008practical} respectively, so the noisy raw $R$ is synthesized as:
\begin{equation}
\label{Eqt.1: noise formation model}
    R \sim K\cdot\mathcal{P}\left(\frac{R^{*}}{K} \right) + \mathcal{N}(0, \sigma_{r}^{2}),
\end{equation}
where $R^{*}$ is the clean raw, $K$ is the total gain related to ISO setting and $\sigma_{r}^{2}$ is the variance of Gaussian noise. $\mathcal{P}(\cdot)$ and $\mathcal{N}(\cdot)$ represent a Poisson and a Gaussian distribution respectively. 

\begin{figure}[t]
  \centering
  \begin{subfigure}{0.55\linewidth}
    \includegraphics[width=\linewidth]{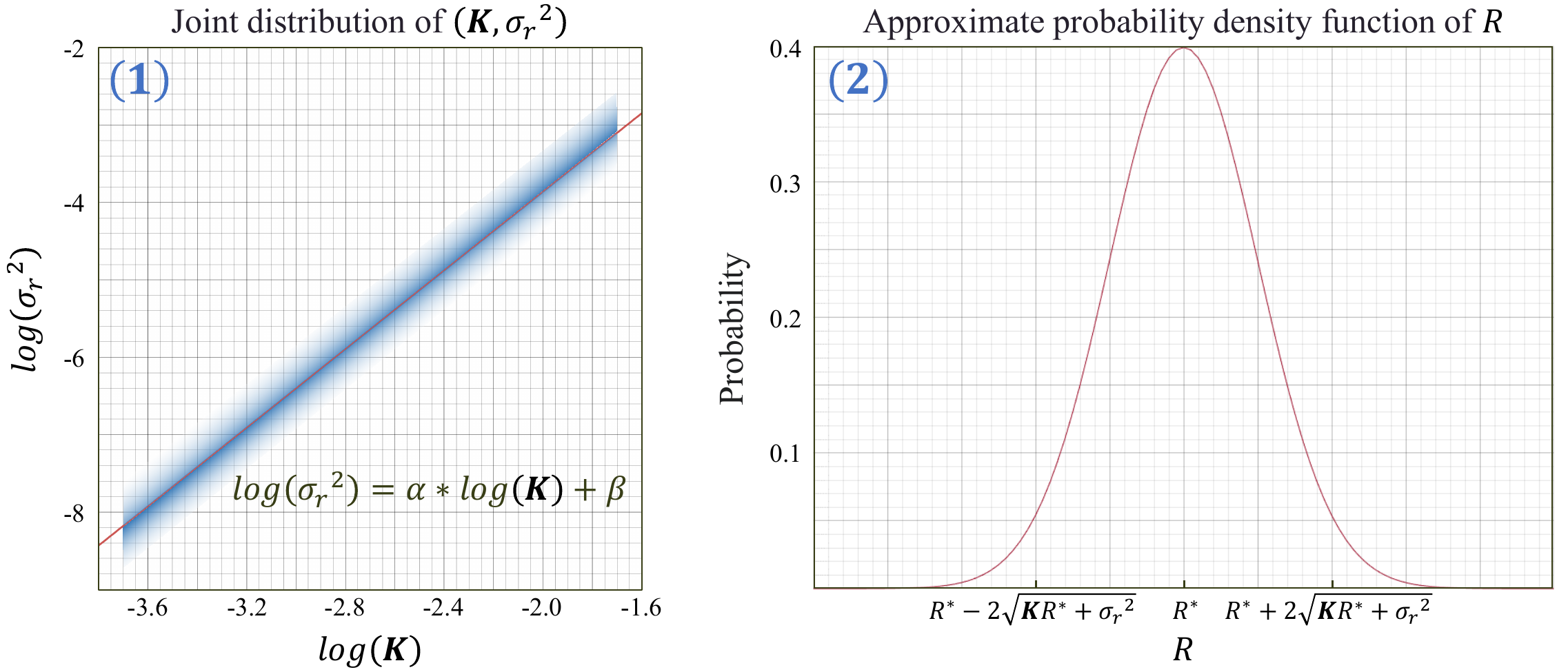}
    \caption{Noise synthesis. (1) We use $K$ to control the generated noise level through the joint distribution. (2) A larger $K$ results in a higher standard deviation of the noise, leading to a more dispersed distribution of $R$.}
    \label{fig:noise_synthesis}
  \end{subfigure}
  \hfill
  \begin{subfigure}{0.4\linewidth}
    \includegraphics[width=\linewidth]{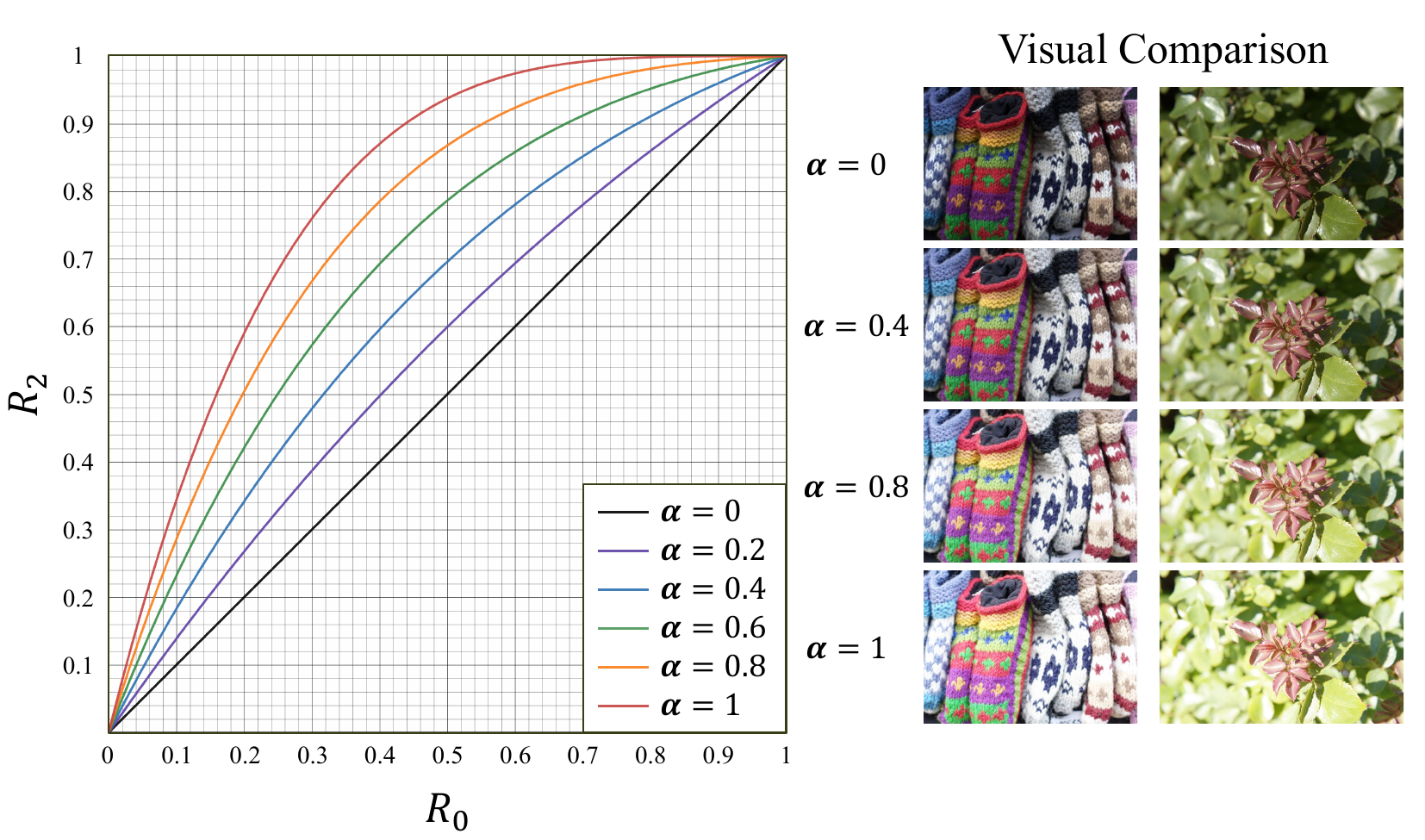}
    \caption{Variable global tone-mapping curve. We use $\alpha$ to control the amplification ratio of global tone-mapping. A larger $\alpha$ results in a brighter sRGB image.}
   \label{fig:variable_TM}
  \end{subfigure}
  \caption{Generate different noise and different ISP amplification ratios during training.}
\end{figure}

To simulate a wide range of noise levels, we model the joint distribution of noise parameters ($K$, $\sigma_{r}^{2}$) and randomly sample from that distribution:
\begin{equation}
\label{Eqt.2: noise parameter sampling procedure}
    \begin{array}{l}
    \log(K) \sim \mathcal{U}\Big(\log(K_{min}), \log(K_{max})\Big),\\
    \log(\sigma_{r}^{2}) \mid \log(K) \sim \mathcal{N}\Big(\alpha \cdot \log(K)+\beta, \sigma^{2}\Big),\\
    \end{array}
\end{equation}
where $\mathcal{U}$ denotes a uniform distribution, $K_{min}$ and $K_{max}$ represent the minimal and maximal noise levels respectively. As shown in~\cref{fig:noise_synthesis}, $\log(\sigma_{r}^{2})$ and $\log(K)$ are linearly correlated with a Gaussian deviation of $\sigma^{2}$, $\alpha$ and $\beta$ indicates the fitted lines slope and intercept. According to~\cite{brooks2019unprocessing,wei2021physics}, the parameters ($\alpha$ = 2.540, $\beta$ = 1.218 and $\sigma$ = 0.268) we adopted here are reasonable in real scenarios. 

Furthermore, we use the cropped patches from full-size images for training. During cropping, we search for the nearest raw patches under the R-G-G-B Bayer pattern around the random location. Therefore, after being processed by pixel-unshuffle operations~\cite{shi2016real}, the resultant packed raw patches have a fixed channel configuration (RGGB), which further improves the model's performance.

\subsection{Differentiable ISP}
\label{sec:Differentiable_ISP}

To train our dual-domain network end-to-end, we designed a differentiable ISP based on PyTorch~\cite{pytorch-paper}.
Although there are some open source ISPs~\cite{simple-camera-pipeline,openISP,karaimer2016software}, they are not differentiable or mutable, limiting the performance and generalization ability of denoising networks if we adopt them. In our framework, we follow a specific order to implement the fundamental parts of ISP, as shown in the right column of~\cref{fig:Our_pipeline}. Unlike most practical ISPs, we exclude certain modules and devise a generalized ISP for two reasons: (1) commercial ISPs are usually black-box functions, and fully simulating them can lead to overfitting; (2) some ISP modules, such as color stylization, have insignificant impact on noise and can be disregarded during training, we select only the necessary modules that affect noise. To test the robustness of our differentiable ISP compared to practical ISPs, we used the well-known RawPy~\cite{rawpy} as a reference. On a set of 50 images, our ISP achieved an average PSNR of 41.7 dB (see the supplement for details).

First of all, we extract the EXIF data from the original raw file, including black level, white balance gain, color correction matrix, and Bayer pattern, and input them to the differentiable ISP. For the Color Filter Array (CFA) interpolation module, We specially modify the classic AHD~\cite{hirakawa2005adaptive} algorithm to mitigate the zipper patterns around edges. For the global tone mapping module, we adopt a mutable curve in the differentiable ISP during training. Since tone mapping may amplify noise and vary in style depending on the desired tuning, we randomly alter the curve instead of fixing it, to more fully simulate the real situation~\cite{brooks2019unprocessing}. The curve should satisfy two constraints: firstly, it should pass through points (0, 0) and (1, 1), and secondly, it should be monotonically increasing with negative second derivatives. Therefore, we use the deep curve~\cite{guo2020zero}, defined as:
\begin{equation}
\label{Eqt.3: Variable Global Tone-mapping Curve}
    R_{n} = R_{n-1} + \alpha\cdot R_{n-1}(1 - R_{n-1}),\\
\end{equation}
where $R_n$ represents the image $R$ after being processed by $n$-th iteration, and $\alpha$ is the amplification ratio that controls the curve shape of tone mapping. Here, we set $n=2$ and randomly sample $\alpha$ from the range [0, 1], and the resultant curves sufficiently cover most cases as visualized in~\cref{fig:variable_TM}.

\subsection{Denoising in Dual-domain}
\label{sec:Denoising_in_Dual_domain}

\noindent\textbf{Noise Map.} 
The noise map was originally designed for raw denoising~\cite{zhang2018ffdnet}, as the noise level highly correlates to the sensor gain. Although the noise map is not new, we are the first ones to use it across raw and sRGB domains. According to ~\cite{foi2008practical}, a Poisson distribution $\mathcal{P}(\frac{R^{*}}{K})$ may be regarded as a special case of Gaussian distribution $\mathcal{N}(\frac{R^{*}}{K}, \frac{R^{*}}{K})$, and thus ~\cref{Eqt.1: noise formation model} can be approximated as follows:
\begin{equation}
\label{Eqt.4: Approximation noise formation model}
    R \sim \mathcal{N}(R^{*}, KR^{*} + \sigma_{r}^{2}).
\end{equation}

We generate noise maps using the standard deviation of clean raw $R^{*}$, namely $\sqrt{KR^{*} + \sigma_{r}^{2}}$, and conduct ablations to verify its effectiveness by comparing different noise maps, including variance, in~\cref{Tab.7: Influence of different noise maps.}. The standard deviation emerges as the best choice to serve as our noise map. It should be noted that the noise-free $R^*$ is not available in practice, so we predict it using the origin $R$.
Additionally, we introduce a Noise map Fusion Block (NFB), which adopts convolutions, concatenation, and a switching mechanism to fuse the input image and the noise map. Detailed information can be found in the supplementary materials.

Previous works have not incorporated noise maps in the sRGB domain, due to the difficulty in accurately predicting the noise level after nonlinear amplification by ISP. However, in our method, we manage to estimate a sRGB noise map for the sRGB denoiser. Since raw denoising will inevitably impair details and cause artifacts, the subsequent sRGB denoising is expected to repair the impairment given the noise map predicting the intensity of impairing. We pass the noise map for the raw denoising through the differentiable ISP, just like the image signal being processed, to obtain the noise map for the sRGB denoising, which fully considers the processing steps prior to entering the sRGB domain. 
Further experiments have confirmed the benefit of using an sRGB noise map in~\cref{sec:ablation_study}.

\noindent\textbf{Dual-denoising Framework.}
Our intention is to plug the dual-denoising networks into a target ISP pipeline, and before the deployment, the dual-denoising networks are trained in the framework as designed in~\cref{fig:Our_pipeline}. In the raw-domain denoising, similar to~\cite{chen2018learning,abdelhamed2019ntire,abdelhamed2020ntire}, each $2\times2$ Bayer pattern is initially packed into four channels using the pixel-unshuffle operation and is unpacked using the pixel-shuffle operation. In the sRGB-domain denoising, the three color channels are mapped directly to the channels of the first and the last layer. As mentioned earlier, the differentiable ISP and noise map mechanisms serve as ``bridges'' between the dual networks, so we adopt long skip connections for denoising networks in both domains, making it feasible to collaboratively train the dual network.

\noindent\textbf{Loss Function.}
We incorporate two stages of supervised learning in the raw and the sRGB domains, using the loss function defined below:
\begin{equation}
\label{Eqt.5: Loss Function}
    L = \lambda\cdot\lVert R-R^{*} \rVert_1 + \lVert I-I^{*} \rVert_1,
\end{equation}
where $\lVert \rVert_1$ represents the $\mathcal{L}_{1}$ loss. $R$, $R^{*}$, $I$, $I^{*}$ are output raw, clean raw, output sRGB, and clean sRGB respectively. Although the differentiable ISP enables our dual-domain model to compute losses and optimize the networks using only sRGB images, ablations in~\cref{tab:ablation_crucial_module} show that adding raw supervision can slightly improve the denoising performance. We adopt $\lambda$ = 1.0 in all experiments.

\section{Experiments}
\label{sec:exp}
In this section, we show different qualitative and quantitative experiments to verify the effectiveness of our approach. As this is the first dual-domain learning-based denoising framework, we will first verify our method outperforms single-domain denoising methods in both visual quality and generalization capability. Then we will also demonstrate that dual-domain denoising can adapt to in-the-wild scenes with severe noise and unseen ISP configurations.

Therefore, we adopt the following evaluation protocol. All training pairs are generated by applying synthetic noise to clean images using the image generation pipeline described in \cref{sec:image_generalization}. For evaluation, we try two different experimental setups: synthetic noise and real noise. For synthetic noise, in addition to using synthetic images, we also experiment with different noise levels, ISP parameters, and even novel ISP modules not present in the training set to demonstrate our generalization ability. For real noise, we test images captured by real cellphones, showing that even when DualDn are trained only on synthetic data and are not fine-tuned on their ISPs or noisy images, it can still effectively remove noise and even outperform commercial denoisers in cellphone cameras.

\subsection{Evaluation on Synthetic Noise}
\label{sec:synthetic_exp}

\myparagraph{Datasets.}
We use the MIT-Adobe FiveK \cite{bychkovsky2011learning} to generate synthetic images for training and testing. It has 5,000 high-resolution raw images, covering a broad range of cameras, scenes, and lighting conditions. It uses DNG raw format with EXIF metadata, which includes ISP parameters to run our differentiable ISP, \eg black level, color matrix, and Bayer pattern \etal. To conduct our evaluation, from the original raw images, we select 220 images captured under low ISO with no observable noise and divide them into 200 training images and 20 test images.

\myparagraph{Backbone Denoisers.}
To demonstrate that our dual-domain denoising is more robust than the single-domain denoising, we choose 3 denoising network structures (SwinIR~\cite{liang2021swinir}, MIRNet-v2~\cite{zamir2022learning}, and Restormer~\cite{zamir2022restormer}) as backbones. Specifically, SwinIR adopts transformer architecture with shifted window attention, MIRNet-v2 performs 2D-convolution to aggregate multi-resolution features, and Restormer uses transformer architecture with channel attention. For a fair comparison, the dual-domain backbone has nearly the same number of parameters as the single-domain one by halving the network blocks.

\begin{figure}[t]
  \centering
   \includegraphics[width=0.925\linewidth]{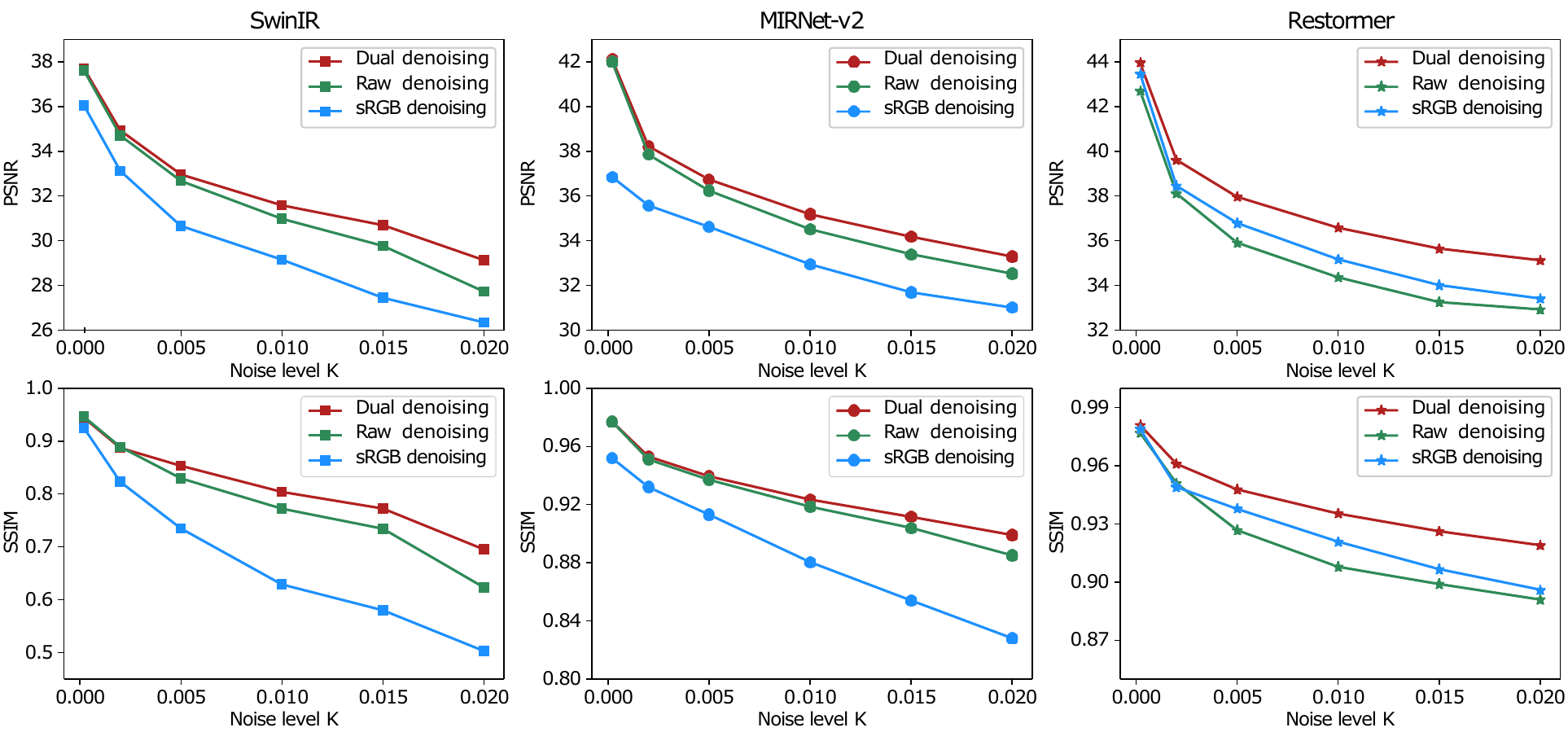}
   \caption{Testing denoising performance with 3 backbones at various noise levels $K$.}
   \label{Fig.: noise level plot}
\end{figure}

\begin{figure}[t]
  \centering
   \includegraphics[width=\linewidth]{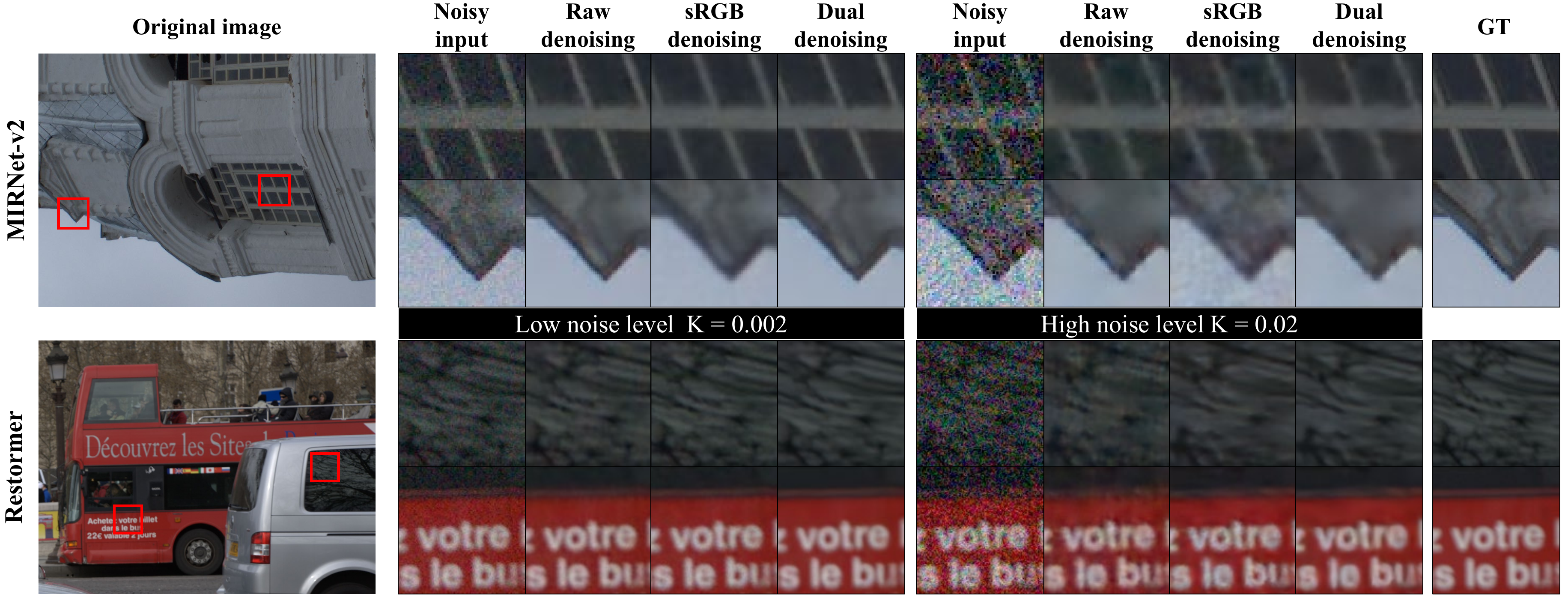}
   \caption{Visual comparisons of denoising results at various noise levels $K$.}
   \label{fig:fixed_alpha}
\end{figure}

\myparagraph{Training Details.}
During training, we randomly crop $256\times256$ patches from the original 4K-resolution raw images, with random horizontal and vertical flips for data augmentation.
All models undergo training on 1 NVIDIA RTX3090 with the AdamW~\cite{loshchilov2017decoupled} optimizer ($\beta_{1}=0.9$, $\beta_{2}=0.999$, $\text{weight decay}=1e^{-2}$) for 120K iterations, and a batch size of 2. The initial learning rate is $1e^{-5}$ and is decreased by a factor of $0.6$ at 40k, 60k, 80k, and 100k iterations.
During testing, to reduce GPU usage, the input raw images are cropped into $512\times512$ patches and the outputs are merged into one image for evaluation. We randomly sample noise parameter $K\sim[0.0002, 0.02]$ and ISP amplification $\alpha\sim[0, 1]$ to increase our model's denoising performance when facing different noise and ISPs.

\myparagraph{Results on Different Noise Levels.} 
First, we evaluate the robustness of our method across various noise levels. We control the noise level through the noise parameter $K$ and set the amplification ratio $\alpha$ to 0 during inference, which eliminates the interference of ISP amplification in this experiment. The results are presented in~\cref{Fig.: noise level plot}. Compared to single-domain denoising, our method achieves consistently better performance across various metrics. Also, the improvement margin increases as the noise becomes stronger (larger $K$), showing that our method possesses a greater ability to generalize to severe noise scenarios. In the first row of~\cref{fig:fixed_alpha}, particularly at $K=0.02$, the color artifacts present in the sRGB denoising are removed through dual denoising. Furthermore, in the second row of~\cref{fig:fixed_alpha}, the zipper noise around the texture edges in the raw denoising is mitigated when switching to dual denoising. Even facing severe noise, DualDn can still achieve visually noise-free results while preserving fine edge details.

\begin{figure}[t]
  \centering
   \includegraphics[width=\linewidth]{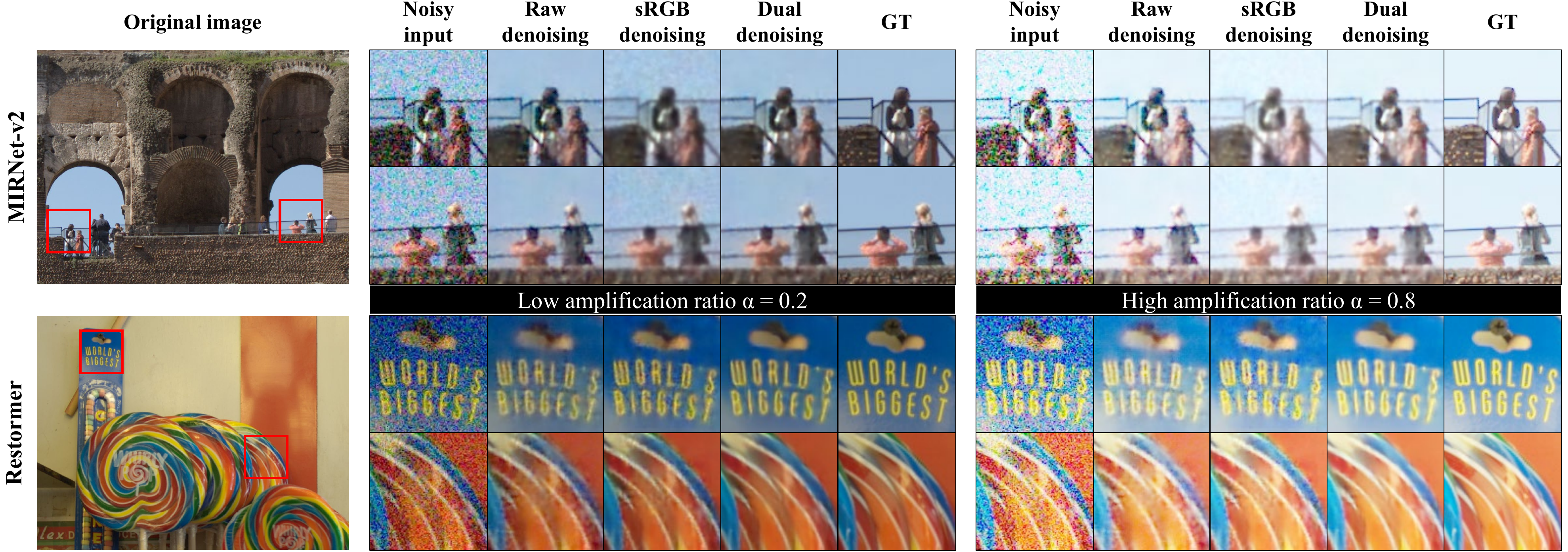}
   \caption{Visual comparisons of denoising results at various ISP amplification ratios $\alpha$. }
   \label{fig:fixed_k}
\end{figure}

\begin{table}[t]
\normalsize

\center{
\caption{Denoising performance at various amplification ratios $\alpha$.}
\label{tab:diff_nonlinear_factor}

\resizebox{\textwidth}{!}{%
\begin{tabular}{llc|ccc|ccc|ccc|cc}
\toprule[1.5pt]
\multicolumn{3}{r}{\textbf{Amplification ratio:}} & \multicolumn{3}{c}{$\boldsymbol{\alpha=0.2}$} & \multicolumn{3}{c}{$\boldsymbol{\alpha=0.5}$} & \multicolumn{3}{c}{$\boldsymbol{\alpha=0.8}$} & Params & Runtime \\ \cmidrule(lr){4-6} \cmidrule(lr){7-9} \cmidrule(lr){10-12}
\multicolumn{2}{l}{\textbf{Backbone}} &  & PSNR$\uparrow$ & SSIM$\uparrow$ & LPIPS$\downarrow$ & PSNR$\uparrow$ & SSIM$\uparrow$ & LPIPS$\downarrow$ & PSNR$\uparrow$ & SSIM$\uparrow$ & LPIPS$\downarrow$ & (G) & (ms)\\ \hline\hline
\multicolumn{2}{l}{\multirow{3}{*}{\textbf{SwinIR}}} & Raw denoising & 26.96 & 0.595 & 0.62 & 25.84 & 0.576 & 0.62 & 24.82 & 0.584 & 0.60 & 11.50 & 45 \\  
\multicolumn{2}{l}{} & sRGB denoising & 26.32 & 0.513 & 0.65 & 25.41 & 0.519 & 0.63 & 23.95 & 0.500 & 0.64 & 11.50 & 212\\ 
\multicolumn{2}{l}{} & Dual denoising (Ours) & \textbf{28.95} & \textbf{0.709} & \textbf{0.50} & \textbf{27.89} & \textbf{0.694} & \textbf{0.50} & \textbf{26.53} & \textbf{0.664} & \textbf{0.50} & 11.79 & 121 \\ \cmidrule(lr){1-14}
\multicolumn{2}{l}{\multirow{3}{*}{\textbf{MIRNet-v2}}} & Raw denoising & 31.47 & 0.865 & \textbf{0.26} & 30.03 & 0.838 & \textbf{0.27} & 28.80 & 0.817 & \textbf{0.27} & 37.48 & 47\\  
\multicolumn{2}{l}{} & sRGB denoising & 30.20 & 0.806 & 0.48 & 28.88 & 0.777 & 0.49 & 27.60 & 0.754 & 0.48 & 37.48 & 55\\ 
\multicolumn{2}{l}{} & Dual denoising (Ours) & \textbf{32.35} & \textbf{0.883} & \textbf{0.26} & \textbf{31.05} & \textbf{0.862} & \textbf{0.27} & \textbf{29.93} & \textbf{0.845} & 0.28 & 38.97 & 54\\ \cmidrule(lr){1-14}
\multicolumn{2}{l}{\multirow{3}{*}{\textbf{Restormer}}} & Raw denoising & 32.08 & 0.873 & 0.23 & 30.65 & 0.850 & 0.24 & 29.44 & 0.831 & 0.25 & 46.23 & 65 \\  
\multicolumn{2}{l}{} & sRGB denoising & 33.01 & 0.889 & \textbf{0.20} & 31.84 & 0.870 & \textbf{0.20} & 30.59 & 0.845 & \textbf{0.20} & 46.23 & 90\\ 
\multicolumn{2}{l}{} & Dual denoising (Ours) & \textbf{33.98} & \textbf{0.906} & 0.22 & \textbf{32.64} & \textbf{0.888} & 0.23 & \textbf{31.48} & \textbf{0.872} & 0.23 & 53.05 & 71\\ \toprule[1.5pt]
\end{tabular}%
}}
\end{table}

\myparagraph{Results on Different ISP Pipelines and Parameters.} One major advantage of our dual-domain approach is its generalization ability to different ISPs, even though it is only trained on a simplified differentiable ISP. To verify that, we evaluate it from two aspects: (1) using the same ISP pipeline but different parameters, and (2) using different ISP pipelines with unseen ISP modules.

\begin{figure}[t]
  \centering
   \includegraphics[width=0.95\linewidth]{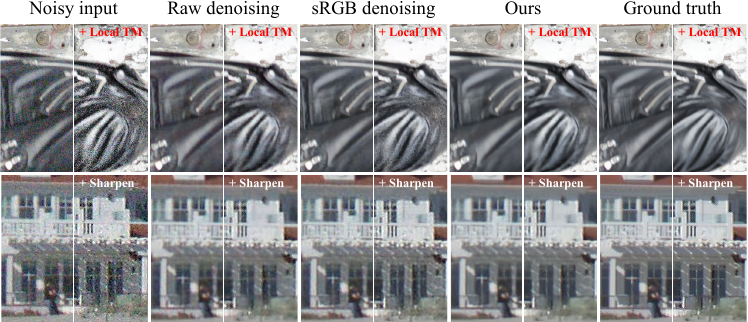}
   \caption{Generalize to unseen ISPs with local tone-mapping (TM) and sharpen modules.}
   \label{fig:unseen_isp_module}
\end{figure}

\begin{table}[t]
\normalsize
\renewcommand\arraystretch{1}
\center{
\caption{Generalization to new ISP pipelines with unseen modules.}
\label{tab:unseen_isp_module}

\resizebox{0.7\textwidth}{!}{%
\begin{tabular}{lllcccccc}
\toprule[1.5pt]
\multicolumn{3}{r}{\textbf{Amplification ratio:}} & \multicolumn{2}{c}{$\boldsymbol{\alpha=0.2}$} & \multicolumn{2}{c}{$\boldsymbol{\alpha=0.5}$} & \multicolumn{2}{c}{$\boldsymbol{\alpha=0.8}$}\\ \cmidrule(lr){4-5} \cmidrule(lr){6-7} \cmidrule(lr){8-9}
\multicolumn{2}{l}{\textbf{Add ISP modules}} & & PSNR & SSIM & PSNR & SSIM & PSNR & SSIM\\ \hline
\multicolumn{2}{c}{} & Raw denoising & 35.50 & 0.928 & 34.09 & 0.913 & 32.91 & 0.899 \\  
\multicolumn{2}{c}{\textbf{+ Sharpen module~~~}} & sRGB denoising & 36.12 & 0.933 & 34.84 & 0.920 & 33.64 & 0.906 \\ 
\multicolumn{2}{c}{} & Dual denoising (Ours) & \textbf{36.90} & \textbf{0.942} & \textbf{35.62} & \textbf{0.929} & \textbf{34.49} & \textbf{0.917} \\ \cmidrule(lr){1-9}
\multicolumn{2}{c}{} & Raw denoising & 23.31 & 0.852 & 23.11 & 0.844 & 23.36 & 0.842 \\  
\multicolumn{2}{l}{\textbf{+ Local TM module~~~}} & sRGB denoising & 22.73 & 0.766 & 22.52 & 0.767 & 22.93 & 0.783 \\ 
\multicolumn{2}{c}{} & Dual denoising (Ours) & \textbf{23.43} & \textbf{0.871} & \textbf{23.25} & \textbf{0.864} & \textbf{23.56} & \textbf{0.862} \\ \toprule[1.5pt]
\end{tabular}%
}}
\end{table}

First, we evaluate the denoising performance with different ISP parameters. We vary the amplification ratios $\alpha$ under the most severe noise ($K = 0.2$) since $\alpha$ directly affects the brightness of the sRGB image, and noise characteristics change maximally at the highest K. Runtime is computed on images with a size of $256\times256$. As shown in~\cref{tab:diff_nonlinear_factor}, across three state-of-the-art backbones, our method significantly improves PSNR and SSIM metrics compared to single denoiser methods at each amplification ratio $\alpha$, with similar parameters and moderate latency. Qualitative comparisons in~\cref{fig:fixed_k} show that while single-domain denoising methods suffer from residual noise and color artifacts, our method retains rich details and is visually closer to the ground truth.

Furthermore, to further demonstrate our denoiser can generalize to unseen ISP pipelines, we select the normal case ($K=0.002$) and insert two additional non-differentiable ISP modules that never appear in training: sharpen and local tone-mapping. The sharpen module is achieved by calculating the high-frequency part and combining it with the original part of the image. Local tone-mapping is implemented through Contrast Limited Adaptive Histogram Equalization (CLAHE)~\cite{zuiderveld1994contrast} to enhance local contrast. As shown in~\cref{fig:unseen_isp_module}, raw denoiser cannot deal with ISP changes and generates worse results after adding ISP modules, sRGB denoiser produces results with more color artifacts because of ISP amplification. Only our method is more tolerant to ISP changes, indicating stronger generalization ability. Qualitative results using Restormer as backbone in~\cref{tab:unseen_isp_module} show our dual domain denoising significantly outperforms baselines.

\begin{figure}[t]
  \centering
   \includegraphics[width=0.93\linewidth]{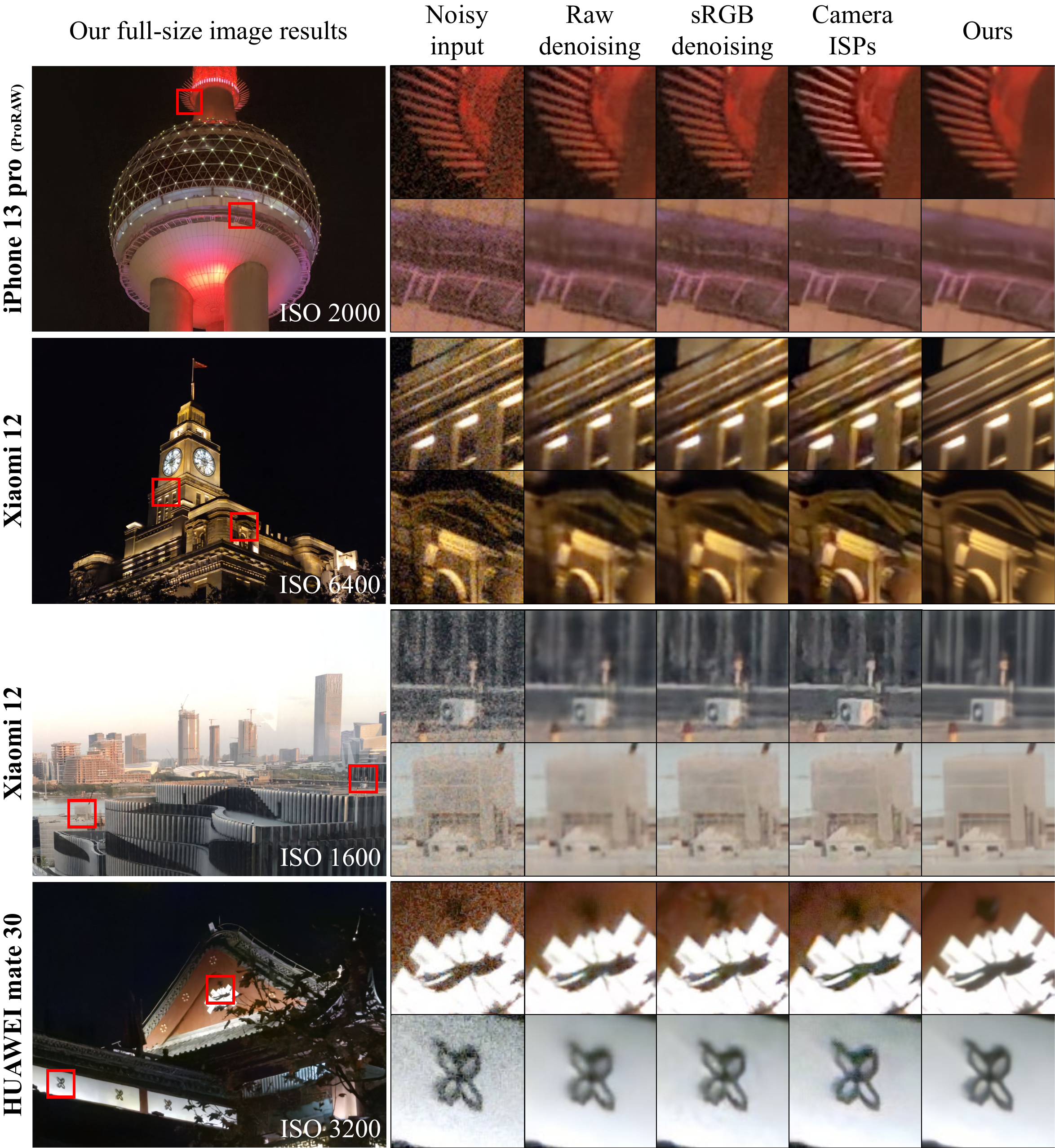}
   \caption{Generalize to unseen smartphone cameras under in-the-wild scenes.}
   \label{Fig.: Visual comparisons of real-world scenarios.}
\end{figure}

\subsection{Evaluation on Real Noise}
\label{sec:Real_world_Scenarios_Testing}
To evaluate the generalization ability to real noisy images, we tested on two datasets. More visual comparisons can be found on our project website.

\myparagraph{In-the-wild Smartphone Evaluation Dataset.}
We first evaluate our algorithms on images captured by flagship mobile devices of 3 well-known brands (Xiaomi, iPhone, and HUAWEI). The commercial ISPs on these devices are normally much more complicated than the differentiable ISPs we used in training, and results on these images show the strong generalization ability of DualDn.

Each device captures 10 different scenes with random ISOs. During inference, we use the pre-trained Restormer model presented in~\cref{sec:synthetic_exp}, without any finetuning on specific devices. As smartphone ISP is a black-box procedure, we can only use ``Pro Mode'' to get raw files with EXIF metadata, so we replace our ISP's variable tone-mapping curve~\cref{Eqt.3: Variable Global Tone-mapping Curve} with bilateral grid upsampling~\cite{chen2016bilateral} to simulate the corresponding ISP for color alignment. \cref{Fig.: Visual comparisons of real-world scenarios.} shows the visual comparison. Even though our model is totally trained on synthetic dataset, it can successfully generalize to real scenes with high-quality denoising, even outperforming the built-in commercial denoising algorithms tailored for specific ISPs. 

\begin{figure}[t]
  \centering
   \includegraphics[width=0.95\linewidth]{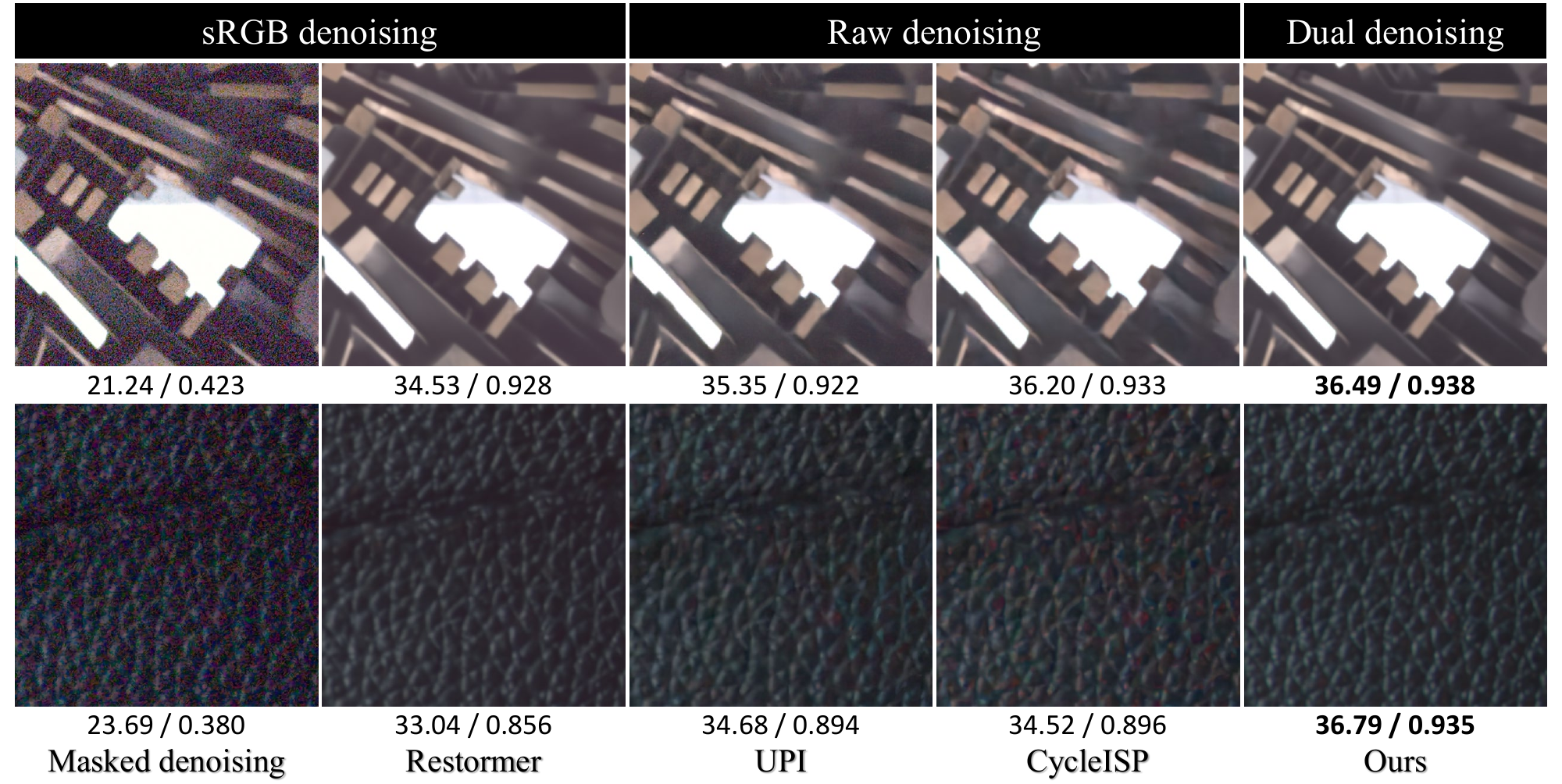}
   \caption{Visual comparisons of DND benchmark results. (PSNR/SSIM)}
   \label{Fig.: DND benchmark.}
\end{figure}

\begin{table}[t]
\small
\renewcommand\arraystretch{1}

\center{
\caption{Raw/sRGB denoising results on the DND benchmark}
\label{tab:Raw/sRGB denoising results on the DND benchmark}

\resizebox{0.6\textwidth}{!}{%
\begin{tabular}{lllcccccc}
\toprule[1pt]
\multicolumn{3}{r}{\textbf{Test domain:}} & \multicolumn{2}{c}{Raw} & \multicolumn{2}{c}{sRGB} \\ \cmidrule(lr){4-5} \cmidrule(lr){6-7} 
\multicolumn{2}{c}{\textbf{Denoising domain}} & & PSNR & SSIM & PSNR & SSIM\\ \hline 
\multicolumn{2}{c}{} & DnCNN~\cite{zhang2017beyond} & 47.37 & 0.976 & 38.08 & 0.936 \\ 
\multicolumn{2}{l}{\textbf{Raw denoising}} & UPI~\cite{brooks2019unprocessing} & 48.88 & 0.982 & 40.35 & 0.964 \\ 
\multicolumn{2}{c}{} & CycleISP~\cite{zamir2020cycleisp} & \underline{49.13} & \textbf{0.983} & \underline{40.50} & \textbf{0.966} \\ 
\multicolumn{2}{c}{} & Ours (intermediate raw output) & \textbf{49.26} & \textbf{0.983} & \textbf{40.70} & \underline{0.964} \\ \cmidrule(lr){1-7}
\multicolumn{2}{c}{} & Masked denoising~\cite{chen2023masked}& - & - & 33.75 & 0.861 \\  
\multicolumn{2}{c}{} & MIRNet~\cite{zamir2020learning} & - & - & 39.72 & 0.959 \\ 
\multicolumn{2}{l}{\textbf{sRGB denoising}} & Uformer~\cite{wang2022uformer} & - & - & 39.77 & 0.959 \\
\multicolumn{2}{c}{} & Restormer~\cite{zamir2022restormer} & - & - & \textbf{40.02} & \textbf{0.960} \\ \cmidrule(lr){1-7} 
\multicolumn{2}{l}{\textbf{Dual denoising}} & Ours & - & - & \textbf{40.59} & \textbf{0.966} \\ \toprule[1pt]
\end{tabular}%
}}
\end{table}

\myparagraph{Public Denoising Benchmark.} Lastly, We validate DualDn on the online benchmark, DND~\cite{plotz2017benchmarking}, consisting of 50 image pairs captured with 4 cameras. Since DND also provides the standard raw with EXIF metadata, we can use it to evaluate our DualDn. Note that DND only uses a fixed and over-simplified ISP with small noise variation, which contradicts the purpose of DualDn designed for complex ISPs with highly varying noise levels. Still, our DualDn achieves comparable results to CycleISP, the best algorithm on the DND dataset, which is specifically designed and fine-tuned for DND's ISP. Here, DualDn adopt Restormer as its backbone and is trained with 5,000 images to match the scale used in~\cite{brooks2019unprocessing,zamir2020cycleisp}.

Specifically, DND~\cite{plotz2017benchmarking} provides both raw and sRGB denoising benchmarks. As shown in~\cref{tab:Raw/sRGB denoising results on the DND benchmark}, our intermediate raw output slightly outperforms state-of-the-art CycleISP. Also, our sRGB results achieve a slightly better PSNR, increasing to 40.59 dB. Unlike the average 1 dB PSNR improvement seen in~\cref{tab:diff_nonlinear_factor}, since DND uses an over-simplified black-box ISP. The discrepancy between our ISP and theirs results in PSNR drops. Moreover, on images with strong noise, our DualDn shows significant improvement over single-domain baselines, as illustrated in~\cref{Fig.: DND benchmark.}. More comparisons in the supplement justify this point.

\subsection{Ablation Study}
\label{sec:ablation_study}

\begin{figure}[t]
  \centering
  \begin{subfigure}{0.5\linewidth}
    \includegraphics[width=\linewidth]{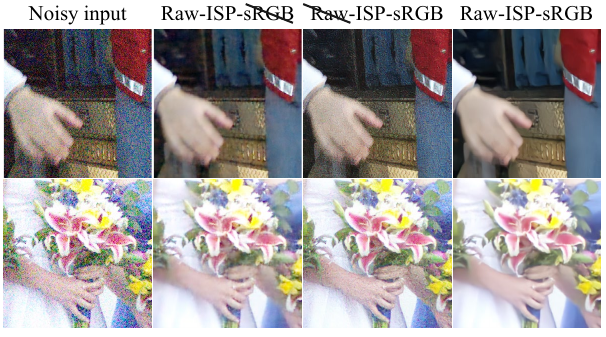}
    \caption{Visual explanation of denoiser functionalities.}
    \label{fig:vis_dual_denoising}
  \end{subfigure}
  \hfill
  \begin{subfigure}{0.48\linewidth}
    \includegraphics[width=\linewidth]{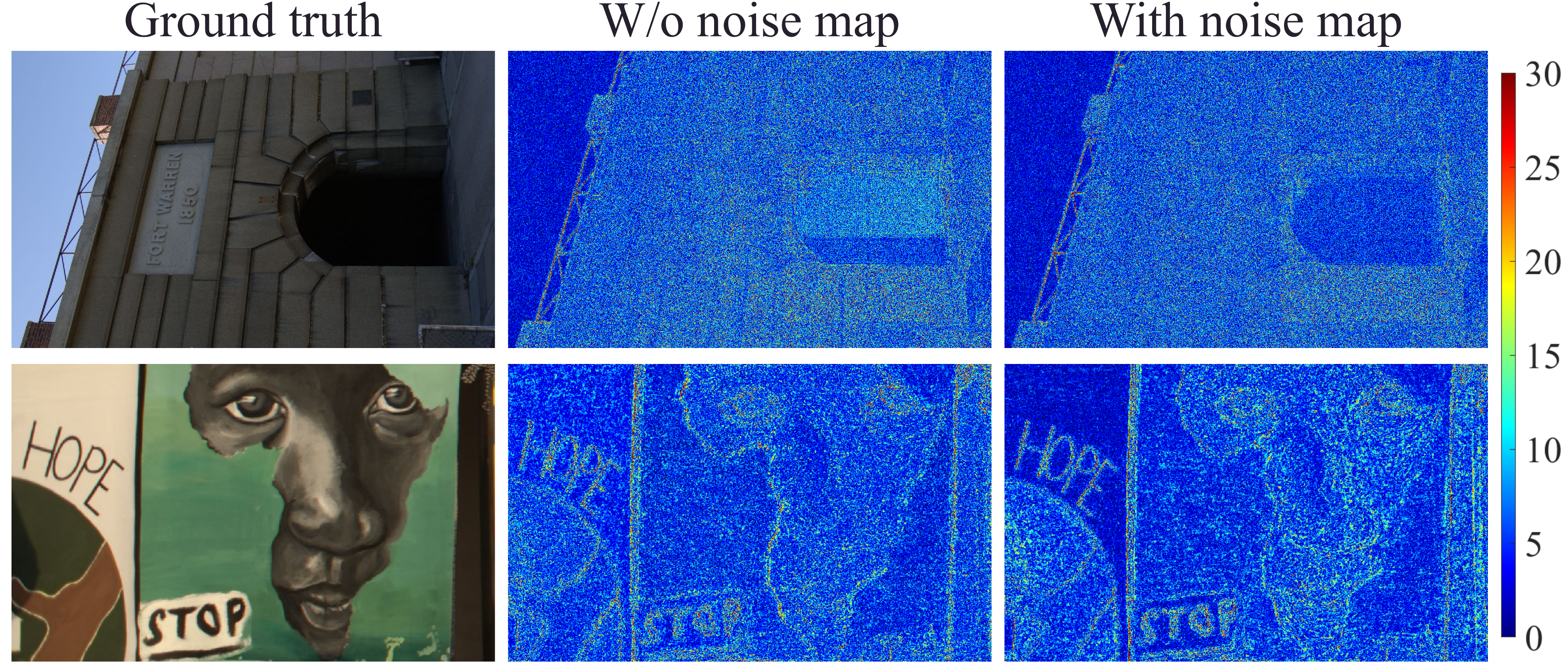}
    \caption{Error maps of the results trained with or w/o noise map. The model trained with noise map shows less difference in the local regions.}
   \label{Fig.6: Noise Map}
  \end{subfigure}
  \caption{Visual explanations of the denoiser functionalities and noise map mechanism.}
\end{figure}

\myparagraph{Functionalities of Dual Denoisers.} 
To illustrate that DualDn can leverage both denoisers by performing preliminary denoising in the raw domain and eliminating residual noise in the sRGB domain, we visualize the results by disabling one of the denoisers during inference. As shown in~\cref{fig:vis_dual_denoising}, the intermediate output after the trained raw denoiser, the ``Raw-ISP'', successfully removes most noise. The remaining noise and artifacts are then eliminated in ``Raw-ISP-sRGB''. An interesting fact is that when testing ``ISP-sRGB'', the trained sRGB denoiser loses its ability to directly denoise the original raw and can only eliminate residual noise after raw denoising, unlike the single sRGB denoiser tested in~\cref{sec:exp}.

\myparagraph{Functionalities of Noise Map.}
Furthermore, we show how noise maps may affect the DualDn's performance. We show a visual comparison of models trained with and without a noise map in~\cref{Fig.6: Noise Map}. Error maps highlight the differences between the results and the ground truth. Models trained with noise maps exhibit significantly smaller errors in local regions. Additionally, we explore the optimal form of noise map under DualDn architecture, as shown in~\cref{Tab.7: Influence of different noise maps.}, the models trained with ``standard deviation'' noise map achieve the best evaluation metrics.

\begin{table}[t]
    \centering
    \begin{subtable}[t]{0.55\linewidth}
    \small
    \resizebox{0.85\textwidth}{!}{
        \begin{tabular}{ccccccccc}
            \toprule[1pt]
            \multicolumn{3}{r}{\textbf{Noise Level:}} & \multicolumn{2}{c}{$\boldsymbol{K=0.0002}$} & \multicolumn{2}{c}{$\boldsymbol{K=0.002}$} & \multicolumn{2}{c}{$\boldsymbol{K=0.02}$}\\ \cmidrule(lr){4-5} \cmidrule(lr){6-7} \cmidrule(lr){8-9}
            \multicolumn{2}{c}{\textbf{Backbone}} & & PSNR & SSIM & PSNR & SSIM & PSNR & SSIM\\ \hline
            \multicolumn{2}{l}{\multirow{4}{*}{\textbf{MIRNet-v2}}} & {w/o noise map} & 38.82 & 0.965 & 36.99 & 0.946 & 32.14 & 0.876 \\  
            \multicolumn{2}{l}{} & {variance} & 40.10 & 0.971 & 37.39 & 0.949 & 31.99 & 0.879 \\ 
            \multicolumn{2}{l}{} & {normalized} & 40.33 & 0.972 & 37.54 & 0.949 & 32.40 & 0.883 \\ 
            \multicolumn{2}{l}{} & {standard deviation} & \textbf{42.01} & \textbf{0.977} & \textbf{37.85} & \textbf{0.951} & \textbf{32.52} & \textbf{0.885} \\ \cmidrule(lr){1-9}
            \multicolumn{2}{l}{\multirow{4}{*}{\textbf{Restormer}}} & {w/o noise map} & 42.46 & 0.977 & 38.05 & 0.950 & 33.14 & 0.891 \\  
            \multicolumn{2}{l}{} & {variance} & 42.20 & 0.976 & 37.86 & 0.949 & 33.01 & 0.887 \\ 
            \multicolumn{2}{l}{} & {normalized} & 42.52 & 0.977 & 37.97 & 0.949 & 32.99 & 0.889 \\ 
            \multicolumn{2}{l}{} & {standard deviation} & \textbf{42.71} & \textbf{0.977} & \textbf{38.12} & \textbf{0.951} & \textbf{33.14} & \textbf{0.891} \\ \toprule[1pt]
        \end{tabular}}
        \caption{Explore the optimal noise map form.}
        \label{Tab.7: Influence of different noise maps.}
    \end{subtable}
    \begin{subtable}[t]{0.4\linewidth}
    \resizebox{\textwidth}{!}{
        \begin{tabular}{cccc|ccc}
            \toprule[1pt]
             Skip & Raw & sRGB & \multirow{2}{*}{Supervision} & \multirow{2}{*}{PSNR} & \multirow{2}{*}{SSIM} \\
             connection & noise map & noise map &  &  &  \\ \toprule
                & \checkmark & \checkmark & $\lambda = 1$ & 34.54 & 0.914 \\ 
             \checkmark &  &  & $\lambda = 1$ & 34.79 & 0.917 \\ 
             \checkmark &  & \checkmark & $\lambda = 1$ & 34.84 & 0.918 \\ 
             \checkmark & \checkmark &  & $\lambda = 1$ & 34.88 & 0.918 \\ 
             \checkmark & \checkmark & \checkmark & $\lambda = 0$ & 34.74 & 0.916 \\ 
             \checkmark & \checkmark & \checkmark & $\lambda = 1$ & \textbf{34.92} & \textbf{0.919} \\ \toprule[1pt]
        \end{tabular}}
        \caption{Ablations of individual modules.}
        \label{tab:ablation_crucial_module}
    \end{subtable}
    \caption{Ablations of our model.}
    \label{tab: Ablations of individual modules.}
\end{table}

\myparagraph{Individual Modules.} At last, we ablate the contributions of different modules in our pipeline, as shown in~\cref{tab:ablation_crucial_module}. In this study, we use $K = 0.02$, $\alpha = 0$. $\lambda$ is a variable used to control raw image supervision in the loss function, as mentioned in~\cref{sec:Denoising_in_Dual_domain}. It is worth noting that even if we train the network using only sRGB images for supervision ($\lambda = 0$), its metrics only drop a little, demonstrating the effectiveness of end-to-end training via differentiable ISP.

\section{Conclusion}
\label{Sec.5: Conclusion}
In this paper, we propose a novel dual-domain denoising method, specifically \textbf{DualDn}, which can effectively remove preliminary noises in the raw domain and the amplified and tangled noises in the sRGB domain caused by ISP processing. The dual-domain network is trained via a differentiable ISP. Nevertheless, the learned denoisers can be directly applied to unknown cameras as plug-and-play modules. Extensive experimental results show that DualDn achieves state-of-the-art denoising performance and generalization ability compared to prior work.

\section*{Acknowledgement}
This work is partially supported by the National Key R\&D Program of China (NO.2022ZD0160201) and CUHK Direct Grants (RCFUS) No. 4055189.
%
%
\bibliographystyle{splncs04}
\bibliography{main}

\begin{thebibliography}{10}
\providecommand{\url}[1]{\texttt{#1}}
\providecommand{\urlprefix}{URL }
\providecommand{\doi}[1]{https://doi.org/#1}

\bibitem{simple-camera-pipeline}
Abdelhamed, A.: Simple camera pipeline. \url{https://github.com/AbdoKamel/simple-camera-pipeline} (2020), [Online; accessed 1-March-2024]

\bibitem{abdelhamed2020ntire}
Abdelhamed, A., Afifi, M., Timofte, R., Brown, M.S.: Ntire 2020 challenge on real image denoising: Dataset, methods and results. In: CVPRW. pp. 496--497 (2020)

\bibitem{abdelhamed2019noise}
Abdelhamed, A., Brubaker, M.A., Brown, M.S.: Noise flow: Noise modeling with conditional normalizing flows. In: CVPR. pp. 3165--3173 (2019)

\bibitem{abdelhamed2018high}
Abdelhamed, A., Lin, S., Brown, M.S.: A high-quality denoising dataset for smartphone cameras. In: CVPR. pp. 1692--1700 (2018)

\bibitem{abdelhamed2019ntire}
Abdelhamed, A., Timofte, R., Brown, M.S.: Ntire 2019 challenge on real image denoising: Methods and results. In: CVPRW. pp.~0--0 (2019)

\bibitem{aharon2006k}
Aharon, M., Elad, M., Bruckstein, A.: K-svd: An algorithm for designing overcomplete dictionaries for sparse representation. IEEE TIP  \textbf{54}(11),  4311--4322 (2006)

\bibitem{boukhayma2016ultra}
Boukhayma, A.: Ultra low noise cmos image sensors. Tech. rep., EPFL (2016)

\bibitem{brooks2019unprocessing}
Brooks, T., Mildenhall, B., Xue, T., Chen, J., Sharlet, D., Barron, J.T.: Unprocessing images for learned raw denoising. In: CVPR. pp. 11036--11045 (2019)

\bibitem{brown2019understanding}
Brown, M., Kim, S.: Understanding color and the in-camera image processing pipeline for computer vision. In: ICCV. Tutorial. pp. 1--247 (2019)

\bibitem{buades2005non}
Buades, A., Coll, B., Morel, J.M.: A non-local algorithm for image denoising. In: CVPR. vol.~2, pp. 60--65. Ieee (2005)

\bibitem{bychkovsky2011learning}
Bychkovsky, V., Paris, S., Chan, E., Durand, F.: Learning photographic global tonal adjustment with a database of input/output image pairs. In: CVPR. pp. 97--104. IEEE (2011)

\bibitem{chan2005salt}
Chan, R.H., Ho, C.W., Nikolova, M.: Salt-and-pepper noise removal by median-type noise detectors and detail-preserving regularization. IEEE Transactions on image processing  \textbf{14}(10),  1479--1485 (2005)

\bibitem{chen2018learning}
Chen, C., Chen, Q., Xu, J., Koltun, V.: Learning to see in the dark. In: CVPR. pp. 3291--3300 (2018)

\bibitem{chen2023masked}
Chen, H., Gu, J., Liu, Y., Magid, S.A., Dong, C., Wang, Q., Pfister, H., Zhu, L.: Masked image training for generalizable deep image denoising. In: CVPR. pp. 1692--1703 (2023)

\bibitem{chen2016bilateral}
Chen, J., Adams, A., Wadhwa, N., Hasinoff, S.W.: Bilateral guided upsampling. ACM TOG  \textbf{35}(6), ~1--8 (2016)

\bibitem{chen2022simple}
Chen, L., Chu, X., Zhang, X., Sun, J.: Simple baselines for image restoration. In: ECCV. pp. 17--33. Springer (2022)

\bibitem{cheng2021nbnet}
Cheng, S., Wang, Y., Huang, H., Liu, D., Fan, H., Liu, S.: Nbnet: Noise basis learning for image denoising with subspace projection. In: CVPR. pp. 4896--4906 (2021)

\bibitem{openISP}
Cruxopen: Open image signal processor. \url{https://github.com/AbdoKamel/simple-camera-pipeline} (2019), [Online; accessed 1-March-2024]

\bibitem{dabov2007image}
Dabov, K., Foi, A., Katkovnik, V., Egiazarian, K.: Image denoising by sparse 3-d transform-domain collaborative filtering. IEEE TIP  \textbf{16}(8),  2080--2095 (2007)

\bibitem{elad2006image}
Elad, M., Aharon, M.: Image denoising via sparse and redundant representations over learned dictionaries. IEEE TIP  \textbf{15}(12),  3736--3745 (2006)

\bibitem{foi2007noise}
Foi, A., Alenius, S., Katkovnik, V., Egiazarian, K.: Noise measurement for raw-data of digital imaging sensors by automatic segmentation of nonuniform targets. IEEE Sensors Journal  \textbf{7}(10),  1456--1461 (2007)

\bibitem{foi2008practical}
Foi, A., Trimeche, M., Katkovnik, V., Egiazarian, K.: Practical poissonian-gaussian noise modeling and fitting for single-image raw-data. IEEE transactions on image processing  \textbf{17}(10),  1737--1754 (2008)

\bibitem{fu2023srgb}
Fu, Z., Guo, L., Wen, B.: srgb real noise synthesizing with neighboring correlation-aware noise model. In: CVPR. pp. 1683--1691 (2023)

\bibitem{goyal2020image}
Goyal, B., Dogra, A., Agrawal, S., Sohi, B.S., Sharma, A.: Image denoising review: From classical to state-of-the-art approaches. Information fusion  \textbf{55},  220--244 (2020)

\bibitem{guo2020zero}
Guo, C., Li, C., Guo, J., Loy, C.C., Hou, J., Kwong, S., Cong, R.: Zero-reference deep curve estimation for low-light image enhancement. In: CVPR. pp. 1780--1789 (2020)

\bibitem{guo2019toward}
Guo, S., Yan, Z., Zhang, K., Zuo, W., Zhang, L.: Toward convolutional blind denoising of real photographs. In: CVPR. pp. 1712--1722 (2019)

\bibitem{henz2020synthesizing}
Henz, B., Gastal, E.S., Oliveira, M.M.: Synthesizing camera noise using generative adversarial networks. IEEE TVCG  \textbf{27}(3),  2123--2135 (2020)

\bibitem{hirakawa2005adaptive}
Hirakawa, K., Parks, T.W.: Adaptive homogeneity-directed demosaicing algorithm. IEEE TIP  \textbf{14}(3),  360--369 (2005)

\bibitem{jiang2022fast}
Jiang, Y., Wronski, B., Mildenhall, B., Barron, J.T., Wang, Z., Xue, T.: Fast and high quality image denoising via malleable convolution. In: ECCV. pp. 429--446. Springer (2022)

\bibitem{jin2023lighting}
Jin, X., Xiao, J.W., Han, L.H., Guo, C., Zhang, R., Liu, X., Li, C.: Lighting every darkness in two pairs: A calibration-free pipeline for raw denoising. In: ICCV. pp. 13275--13284 (2023)

\bibitem{karaimer2016software}
Karaimer, H.C., Brown, M.S.: A software platform for manipulating the camera imaging pipeline. In: ECCV. pp. 429--444. Springer (2016)

\bibitem{kingma2018glow}
Kingma, D.P., Dhariwal, P.: Glow: Generative flow with invertible 1x1 convolutions. Advances in neural information processing systems  \textbf{31} (2018)

\bibitem{knaus2013dual}
Knaus, C., Zwicker, M.: Dual-domain image denoising. In: ICIP. pp. 440--444. IEEE (2013)

\bibitem{konnik2014high}
Konnik, M., Welsh, J.: High-level numerical simulations of noise in ccd and cmos photosensors: review and tutorial. arXiv preprint arXiv:1412.4031  (2014)

\bibitem{rawpy}
Letmaik: Rawpy. \url{https://github.com/letmaik/rawpy}, [Online; accessed 1-March-2024]

\bibitem{li2023ntire}
Li, Y., Zhang, Y., Timofte, R., Van~Gool, L., Tu, Z., Du, K., Wang, H., Chen, H., Li, W., Wang, X., et~al.: Ntire 2023 challenge on image denoising: Methods and results. In: CVPRW. pp. 1904--1920 (2023)

\bibitem{liang2015stacked}
Liang, J., Liu, R.: Stacked denoising autoencoder and dropout together to prevent overfitting in deep neural network. In: 2015 8th international congress on image and signal processing (CISP). pp. 697--701. IEEE (2015)

\bibitem{liang2021swinir}
Liang, J., Cao, J., Sun, G., Zhang, K., Van~Gool, L., Timofte, R.: Swinir: Image restoration using swin transformer. In: ICCV. pp. 1833--1844 (2021)

\bibitem{liang2021mutual}
Liang, J., Sun, G., Zhang, K., Van~Gool, L., Timofte, R.: Mutual affine network for spatially variant kernel estimation in blind image super-resolution. In: CVPR. pp. 4096--4105 (2021)

\bibitem{loshchilov2017decoupled}
Loshchilov, I., Hutter, F.: Decoupled weight decay regularization. arXiv preprint arXiv:1711.05101  (2017)

\bibitem{mairal2009non}
Mairal, J., Bach, F., Ponce, J., Sapiro, G., Zisserman, A.: Non-local sparse models for image restoration. In: ICCV. pp. 2272--2279. IEEE (2009)

\bibitem{monakhova2022dancing}
Monakhova, K., Richter, S.R., Waller, L., Koltun, V.: Dancing under the stars: video denoising in starlight. In: CVPR. pp. 16241--16251 (2022)

\bibitem{pytorch-paper}
Paszke, A., Gross, S., Chintala, S., Chanan, G., Yang, E., DeVito, Z., Lin, Z., Desmaison, A., Antiga, L., Lerer, A.: Automatic differentiation in pytorch  (2017), \url{https://pytorch.org}

\bibitem{plotz2017benchmarking}
Plotz, T., Roth, S.: Benchmarking denoising algorithms with real photographs. In: CVPR. pp. 1586--1595 (2017)

\bibitem{shi2016real}
Shi, W., Caballero, J., Husz{\'a}r, F., Totz, J., Aitken, A.P., Bishop, R., Rueckert, D., Wang, Z.: Real-time single image and video super-resolution using an efficient sub-pixel convolutional neural network. In: CVPR. pp. 1874--1883 (2016)

\bibitem{tian2020deep}
Tian, C., Fei, L., Zheng, W., Xu, Y., Zuo, W., Lin, C.W.: Deep learning on image denoising: An overview. Neural Networks  \textbf{131},  251--275 (2020)

\bibitem{wang2022uformer}
Wang, Z., Cun, X., Bao, J., Zhou, W., Liu, J., Li, H.: Uformer: A general u-shaped transformer for image restoration. In: CVPR. pp. 17683--17693 (2022)

\bibitem{wei2020physics}
Wei, K., Fu, Y., Yang, J., Huang, H.: A physics-based noise formation model for extreme low-light raw denoising. In: CVPR. pp. 2758--2767 (2020)

\bibitem{wei2021physics}
Wei, K., Fu, Y., Zheng, Y., Yang, J.: Physics-based noise modeling for extreme low-light photography. IEEE TPAMI  \textbf{44}(11),  8520--8537 (2021)

\bibitem{xu2015denoising}
Xu, Q., Zhang, C., Zhang, L.: Denoising convolutional neural network. In: 2015 IEEE International Conference on Information and Automation. pp. 1184--1187. IEEE (2015)

\bibitem{yapici2021review}
Yapici, A., Akcayol, M.A.: A review of image denoising with deep learning. In: 2021 2nd International Informatics and Software Engineering Conference (IISEC). pp.~1--6. IEEE (2021)

\bibitem{zamir2022restormer}
Zamir, S.W., Arora, A., Khan, S., Hayat, M., Khan, F.S., Yang, M.H.: Restormer: Efficient transformer for high-resolution image restoration. In: CVPR. pp. 5728--5739 (2022)

\bibitem{zamir2020cycleisp}
Zamir, S.W., Arora, A., Khan, S., Hayat, M., Khan, F.S., Yang, M.H., Shao, L.: Cycleisp: Real image restoration via improved data synthesis. In: CVPR. pp. 2696--2705 (2020)

\bibitem{zamir2020learning}
Zamir, S.W., Arora, A., Khan, S., Hayat, M., Khan, F.S., Yang, M.H., Shao, L.: Learning enriched features for real image restoration and enhancement. In: ECCV. pp. 492--511. Springer (2020)

\bibitem{zamir2022learning}
Zamir, S.W., Arora, A., Khan, S., Hayat, M., Khan, F.S., Yang, M.H., Shao, L.: Learning enriched features for fast image restoration and enhancement. IEEE TPAMI  \textbf{45}(2),  1934--1948 (2022)

\bibitem{zhang2017improved}
Zhang, J., Hirakawa, K.: Improved denoising via poisson mixture modeling of image sensor noise. IEEE TIP  \textbf{26}(4),  1565--1578 (2017)

\bibitem{zhang2021plug}
Zhang, K., Li, Y., Zuo, W., Zhang, L., Van~Gool, L., Timofte, R.: Plug-and-play image restoration with deep denoiser prior. IEEE TPAMI  \textbf{44}(10),  6360--6376 (2021)

\bibitem{zhang2021designing}
Zhang, K., Liang, J., Van~Gool, L., Timofte, R.: Designing a practical degradation model for deep blind image super-resolution. In: CVPR. pp. 4791--4800 (2021)

\bibitem{zhang2017beyond}
Zhang, K., Zuo, W., Chen, Y., Meng, D., Zhang, L.: Beyond a gaussian denoiser: Residual learning of deep cnn for image denoising. IEEE TIP  \textbf{26}(7),  3142--3155 (2017)

\bibitem{zhang2018ffdnet}
Zhang, K., Zuo, W., Zhang, L.: Ffdnet: Toward a fast and flexible solution for cnn-based image denoising. IEEE TIP  \textbf{27}(9),  4608--4622 (2018)

\bibitem{zhang2023real}
Zhang, Z., Jiang, Y., Shao, W., Wang, X., Luo, P., Lin, K., Gu, J.: Real-time controllable denoising for image and video. In: CVPR. pp. 14028--14038 (2023)

\bibitem{zhu2016noise}
Zhu, F., Chen, G., Heng, P.A.: From noise modeling to blind image denoising. In: CVPR. pp. 420--429 (2016)

\bibitem{zoran2011learning}
Zoran, D., Weiss, Y.: From learning models of natural image patches to whole image restoration. In: ICCV. pp. 479--486. IEEE (2011)

\bibitem{zuiderveld1994contrast}
Zuiderveld, K.: Contrast limited adaptive histogram equalization. Graphics gems pp. 474--485 (1994)

\end{thebibliography}
\end{document}